\documentclass[%
reprint,
showpacs,preprintnumbers,
amsmath,amssymb,
aps,
prc,
]{revtex4-1}
\usepackage{graphicx}
\usepackage{bm}
\usepackage{epstopdf}

\usepackage{amsmath}
\usepackage{bm}
\usepackage{mathtools}
\usepackage{siunitx}
\DeclareSIUnit\clight{\text{\ensuremath{c}}}
\DeclareSIUnit\fm{\femto\meter}
\newcommand{\Y}{\mathrm{\Upsilon}}
\newcommand{\YnS}[1]{{\Y({#1}\mathrm{S})}}
\newcommand{\chib}{\mathrm{\chi_b}}
\newcommand{\chibnP}[1]{{\chib({#1}\mathrm{P})}}

\newcommand{\ncoll}{{\langle n_\text{coll} \rangle}}
\newcommand{\Ncoll}{{\langle N_\text{coll} \rangle}}
\newcommand{\Npart}{{\langle N_\text{part} \rangle}}

\newcommand{\sqrtsNN}{\sqrt{\smash[b]{s_\text{NN}}}}
\newcommand{\tauF}{{\tau_\text{F}}}
\newcommand{\Tcrit}{{T_\text{c}}}
\newcommand{\Tinit}{{T_0}}

\begin{document}

\preprint{APS/123-QED}
\title{Hot-medium effects on $\mathbf{\Upsilon}$~yields in pPb~collisions at $\sqrt{\smash[b]{\text{\textit{s}\textsubscript{NN}}}} = \SI{8.16}{\TeV}$}
\author{V.\,H.\,Dinh}
\author{J.\,Hoelck}
\author{G.\,Wolschin}
\email{g.wolschin@thphys.uni-heidelberg.de}
\affiliation{Institut f{\"ur} Theoretische Physik der Universit{\"a}t Heidelberg, Philosophenweg 12-16, D-69120 Heidelberg, Germany, EU}
\date{\today}

\begin{abstract}
	The respective contributions of cold-matter and hot-medium effects to the suppression of $\YnS{1}$ and $\YnS{2}$~mesons in pPb collisions at energies reached at the Large Hadron Collider (LHC) are investigated.
	Whereas known alterations of the parton density functions in the lead nucleus and coherent parton energy loss
	account for the leading fraction of the modifications in cold nuclear matter (CNM), the hot-medium (quark-gluon plasma, QGP) effects turn out to be relevant in spite of the small initial spatial extent of the fireball.
	We compare our transverse-momentum-, rapidity-, and centrality-dependent theoretical results for the $\Y$ suppression in pPb collisions at a center-of-mass energy of $\sqrtsNN = \SI{8.16}{\TeV}$ with recent LHCb and preliminary ALICE data from the Large Hadron Collider (LHC).
	Both cold-matter and hot-medium effects are needed to account for the data.
	The initial central temperature of the fireball is found to be $\Tinit \simeq \SI{460}{\MeV}$.
\end{abstract}

\pacs{25.75.-q,25.75.Dw,25.75.C}
\maketitle

\section{Introduction}
\label{sec:intro}
The successive suppression of the bottomonia states $\YnS{1}$, $\YnS{2}$, and $\YnS{3}$ in the hot quark-gluon plasma (QGP) that is created in symmetric high-energy heavy-ion collisions at the Relativistic Heavy Ion Collider (RHIC) \cite{ada14} and the Large Hadron Collider (LHC) \cite{CMS-2012,ab14,cms19,alice19} is a valuable indicator for its properties, such as the initial central temperature $\Tinit$ \cite{brezinski-wolschin-2012,em12,striba12,peng11,son12,ngw13,ngw14,hnw17}.
In smaller asymmetric systems like pPb, however, the fireball with temperatures exceeding the critical value $T=\Tcrit \simeq \SI{160}{\MeV}$ is spatially much less extended than in PbPb or AuAu.
Correspondingly, cold nuclear matter (CNM) effects contribute significantly to the modification of bottomonia yields in asymmetric collisions when compared to pp.

In this work, we explore the contribution of both cold-matter and hot-medium effects on the modification of bottomonia yields in pPb collisions at $\sqrtsNN = \SI{8.16}{\TeV}$.
Transverse-momentum-, rapidity- and centrality-dependent results are compared with recent CMS \cite{cms19} and preliminary ALICE \cite{alice19} data.

Regarding the CNM-effects, we consider the modification of the parton distribution functions (PDFs) in a nucleus compared to free nucleons, and the coherent energy loss of the bottomonia on their paths through the medium \cite{alba18}.
These cold-matter effects are not expected to be much different for ground and excited bottomonia states; indeed, the ALICE measurement of the cross section ratio $\YnS{2}/\YnS{1}$ in \SI{5.02}{\TeV} pPb shows no evidence for different CNM effects on the two states at forward and backward rapidities, albeit within large uncertainties \cite{alice15}.

For the hot-medium effects, we rely on our model for bottomonia suppression in the QGP that we have developed for heavy systems such as AuAu and PbPb, but we adapt it now to the case of small asymmetric systems.
The model is based on gluon-induced dissociation, screening of the real part of the quark-antiquark potential, and damping through the imaginary part \cite{ngw13,ngw14,hnw17}.
A significant fraction of the $\YnS{1}$ suppression can also be due to the reduced feed-down from excited states, once these are mostly screened, or depopulated, as is the case in heavy systems at RHIC and LHC energies.
In particular for \SI{5.02}{\TeV} PbPb collisions at the LHC, the model has proven its predictive properties \cite{gw19} regarding the transverse-momentum and centrality dependence of the $\Y$ suppression.

Although the spatial overlap of projectile and target in pPb, and hence the initial QGP region before hydrodynamic expansion and gradual cooling, is significantly smaller in pPb, we conjecture that the basic suppression mechanisms in the hot medium remain unaltered.
The emphasis in this work will then be on the interplay of cold-matter and hot-medium effects, and their comparison to
data for \SI{8.16}{\TeV} pPb collisions at the LHC.

In Sec.\,\ref{sec:pops}, we consider the initial populations of the bottomonia states.
Aspects of the treatment of CNM effects in \SI{8.16}{\TeV} pPb collisions are briefly reviewed in Sec.\,\ref{sec:cnm}.
These had extensively been considered by a large group of authors in Ref.\,\cite{alba18}.
For the initial-state modification of the parton distribution functions in the nuclear medium, we use the most recent global analysis of nuclear shadowing that provides a new set of PDFs, Ref.\,\cite{esk17}.
It includes, in particular, LHC data from the \SI{5.02}{\TeV} pPb run.
Coherent parton energy loss is also accounted for, in the model of Arleo and Peign{\'e} \cite{arleo13,arl13,arl16}.
In Sec.\,\ref{sec:qgp}, we consider the hot-medium effects in our model \cite{hnw17} that we have originally developed for symmetric heavy systems such as PbPb.
We reconsider the main aspects of the model, tailoring it now to the case of smaller and asymmetric systems such as pPb.
The resulting
calculations and comparisons with data from two LHC-collaborations obtained in the \SI{8.16}{\TeV} pPb run
are presented in Sec.\,\ref{sec:results}, the conclusions are drawn in Sec.\,\ref{sec:conclusion}.

\section{Bottomonia populations}
\label{sec:pops}
\begin{figure}
	\centering
	\includegraphics{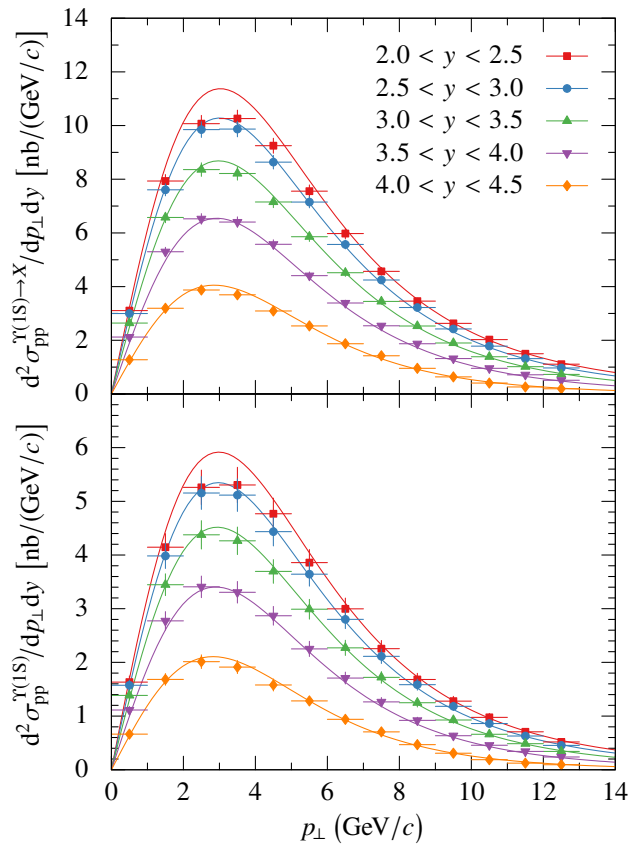}
	\caption{\label{fig1}
		(color online)
		Fits of the double-differential cross sections for $\YnS{1}$ decays (top) and $\YnS{1}$ production (bottom) in $\sqrt{s}=\SI{8}{\TeV}$ pp collisions.
		The data points and error bars are based on LHCb data \cite{lhcb15}.
		Our corresponding fits are displayed as functions of transverse momentum for five rapidity regions.
	}
\end{figure}

The production of $\Y$ mesons in proton-proton collisions can occur either directly in parton scattering, or via feed-down from the decay of heavier bottomonium states, such as $\chib$, or higher-mass $\Y$ states, thus complicating the theoretical description of bottomonium production.
In this work, we make use of the experimentally measured double-differential pp cross sections at \SI{8}{\TeV} of dimuon pairs from $\Y$ decays, $\mathrm{d}^2\sigma_\text{pp}^{\Y\to\mathrm{\mu^+\mu^-}}\!/(\mathrm{d}p_\perp\mathrm{d}y)$, from the LHCb collaboration \cite{lhcb15}.
These data are rescaled using the corresponding dimuon branching ratios to obtain the inclusive bottomonium-decay cross sections in pp collisions, $\mathrm{d}^2\sigma_\text{pp}^{\Y\to X}\!/(\mathrm{d}p_\perp\mathrm{d}y)$.
Then, we apply an inverse feed-down cascade \cite{ngw13,vnw13,hnw17} for every $p_\perp$ and $y$ bin to reconstruct the (direct) bottomonium-production cross sections in pp collisions, $\mathrm{d}^2\sigma_\text{pp}^\Y/(\mathrm{d}p_\perp\mathrm{d}y)$.
The latter do not include the indirect contributions from feed-down and hence, are always smaller than the measured decay cross sections.
We fit decay and production cross sections separately with an analytical fit function proposed in Ref.\,\cite{arl13},
\begin{align}
	\frac{\mathrm{d}^2\sigma_\text{pp}}{\mathrm{d}p_\perp\mathrm{d}y}
	= \mathcal{N} \, p_\perp \left(\frac{p_0^2}{p_0^2 + p_\perp^2}\right)^{\!m} \left(1 - \frac{2M_\perp}{\sqrt{s}} \cosh y\right)^{\!n}.
	\label{eq:fitfunction}
\end{align}
The errors of the decay cross sections are given by the experimental uncertainties.
For the bottomonium-production cross sections, we propagate the errors of the dimuon data, which we assume to be uncorrelated, through the inverse decay cascade.
The fits are shown in Fig.\,\ref{fig1}.
Apart from the smallest rapidity bin, the fits are sufficiently precise, with an overall $\chi^2/\mathrm{ndf} = 2.12$ and $\chi^2/\mathrm{ndf} = 0.50$ for the $\YnS{1}$-decay and $\YnS{1}$-production cross sections, respectively, cf.~Tab.\,\ref{tab:Fit}.
Hence, we build the subsequent calculations for the $\YnS{1}$ and, similarly, $\YnS{2}$ yields in pPb collisions on the analytical functions.

\begin{table}[!h]
	\centering
	\caption{\label{tab:Fit}
		Fit parameters of the double-differential cross sections for $\YnS{1}$ decays and $\YnS{1}$ production in $\sqrt{s} = \SI{8}{\TeV}$ pp collisions.
	}
	\begin{ruledtabular}
		\begin{tabular}{l *{5}{c}}
			& $\mathcal{N}$ & $p_0$ & $m$ & $n$ & $\chi^2/\mathrm{ndf}$ \\
			\colrule
			decays & 7.61 & 6.18 & 2.55 & 13.36 & 2.12 \\
			production & 4.00 & 6.23 & 2.64 & 13.38 & 0.50 \\
		\end{tabular}
	\end{ruledtabular}
\end{table}

\section{Cold-matter effects}
\label{sec:cnm}
In asymmetric collisions such as pPb, the bottomonia yields are already influenced by the presence of nuclear matter -- even if a hot medium was completely absent.
These well-known cold nuclear matter effects include purely initial-state effects -- in particular, the modification of the initial gluon densities --, as well as mixed initial- and final-state effects such as the coherent parton energy loss induced by the nuclear medium.
We subsequently consider the above two CNM-effects and determine the corresponding modification of bottomonia yields in pPb as compared to what is expected from pp collisions at LHC energies.

\begin{table}
	\centering
	\caption{\label{tab:Glauber}
		Results of our Glauber calculation for the expected numbers of binary collisions~$\Ncoll$ and participants~$\Npart$, the differential and integrated inelastic pPb cross sections $\mathrm{d}\sigma_\text{pPb}^\text{inel}/\mathrm{d}b$ and $\sigma_\text{pPb}^\text{inel}$, and the corresponding centrality~$c$ in pPb collisions at $\sqrtsNN=\SI{8.16}{\TeV}$ for different impact parameters~$b$.
	}
	\begin{ruledtabular}
		\begin{tabular}{*{6}{r}}
			{$b$} & {$\Ncoll$} & {$\Npart$} & {$\mathrm{d}\sigma_\text{pPb}^\text{inel}/\mathrm{d}b$} & {$\sigma_\text{pPb}^\text{inel}$} & {$c$} \\
			{(\si{\fm})} & & & {(\si{\fm})} & {(\si{\fm\squared})} & {(\si{\percent})} \\
			\colrule
			 0 & 15.6 & 16.6 &    0 &   0 &    0 \\
			 1 & 15.4 & 16.4 &  6.3 &   3 &  1.6 \\
			 2 &  4.8 & 15.8 & 12.6 &  13 &  6.0 \\
			 3 &  3.7 & 14.7 & 18.8 &  29 & 13.3 \\
			 4 &  2.0 & 12.9 & 25.1 &  52 & 23.5 \\
			 5 &  9.3 & 10.3 & 31.4 &  80 & 36.6 \\
			 6 &  6.0 &  6.9 & 37.6 & 115 & 52.4 \\
			 7 &  2.8 &  3.6 & 41.4 & 155 & 70.8 \\
			 8 &  0.9 &  1.4 & 30.6 & 192 & 87.7 \\
			 9 &  0.2 &  0.4 & 11.4 & 212 & 96.7 \\
			10 &  0.0 &  0.1 &  2.7 & 218 & 99.3 \\
			11 &  0.0 &  0.0 &  0.5 & 219 & 99.9 \\
		\end{tabular}
	\end{ruledtabular}
\end{table}

\subsection{Modification of Bottomonium Production}
The modification from pp to pPb collisions is quantified by the nuclear modification factor
\begin{align}
	R_\text{pPb}(b, p_\perp, y) = \frac{1}{\Ncoll(b)}\frac{\frac{\mathrm{d}^2\sigma_\text{pPb}^{\Y\to X}}{\mathrm{d}p_\perp\mathrm{d}y}(b,p_\perp,y)}{\frac{\mathrm{d}^2\sigma_\text{pp}^{\Y\to X}}{\mathrm{d}p_\perp\mathrm{d}y}(p_\perp,y)}~,
\end{align}
where $\mathrm{d}^2\sigma^{\Y\to X}\!/(\mathrm{d}p_\perp\mathrm{d}y)$ is the Lorentz-invariant double-differential cross section for $\Y$~decays.
For pPb, these cross sections are calculated via the decay cascade~\cite{vnw13} from the corresponding production cross sections after applying all cold-matter (and later, hot-medium) modifications.
$R_\text{pPb}$ depends on the rapidity~$y$, transverse momentum~$p_\perp$, and the centrality of the collision which can be expressed either in terms of the impact parameter~$b$ or by the expected number of binary nucleon-nucleon collisions~$\Ncoll(b)$, see Tab.\,\ref{tab:Glauber}.
Modification factors as functions of only one observable are obtained by integrating the differential cross sections first before taking the ratio,
\begin{align}
	\label{eq:cFactor}
	R_\text{pPb}(b) &= \frac{1}{\Ncoll}\frac{\iint\frac{\mathrm{d}^2\sigma_\text{pPb}^{\Y\to X}}{\mathrm{d}p_\perp\mathrm{d}y}\mathrm{d}p_\perp\mathrm{d}y}{\iint\frac{\mathrm{d}^2\sigma_\text{pp}^{\Y\to X}}{\mathrm{d}p_\perp\mathrm{d}y}\mathrm{d}p_\perp\mathrm{d}y}\,,\\
	\label{eq:pTFactor}
	R_\text{pPb}(p_\perp) &= \frac{\iint\frac{\mathrm{d}^2\sigma_\text{pPb}^{\Y\to X}}{\mathrm{d}p_\perp\mathrm{d}y}\frac{\mathrm{d}\sigma_\text{pPb}^\text{inel}}{\mathrm{d}b}\mathrm{d}b\,\mathrm{d}y}{\int\Ncoll\frac{\mathrm{d}\sigma_\text{pPb}^\text{inel}}{\mathrm{d}b}\mathrm{d}b \, \int\frac{\mathrm{d}^2\sigma_\text{pp}^{\Y\to X}}{\mathrm{d}p_\perp\mathrm{d}y}\mathrm{d}y}\,,
\end{align}
and $R_\text{pPb}(y)$ accordingly.
Here, ${\mathrm{d}\sigma_\text{pPb}^\text{inel}}/{\mathrm{d}b}$ is the differential inelastic pPb cross section which is given by the Glauber model \cite{Biallas1976}.
We determine the bottomonium-production cross sections in pPb collisions by adjusting the corresponding pp cross sections which we parameterized with Eq.\,(\ref{eq:fitfunction}).

For our calculations, we consider the nuclear gluon shadowing and the coherent parton energy loss model to account for the CNM effects.
Thus, the CNM modification of the bottomonium-production cross section from pp to pPb collisions reads \cite{arl13, vogt15}
\begin{widetext}
\begin{equation}
	\label{eq:AllEffects}
	\frac{1}{\Ncoll} \frac{\mathrm{d}^2\sigma_\text{pPb}^\text{CNM}}{\mathrm{d}p_\perp \mathrm{d}y} = \int_0^{2 \pi} \frac{\mathrm{d}\varphi}{2 \pi} \int_0^{\varepsilon_\text{max}} \mathrm{d}\varepsilon \,
	P(\varepsilon, E, L_\text{eff}) \,
	\frac{p_\parallel}{p_\parallel^\text{shift}}\frac{p_\perp}{p_\perp^\text{shift}} \,
	R_g^\text{Pb}(x_2^\text{shift}) \,
	\frac{\mathrm{d}^2\sigma_\text{pp}}{\mathrm{d}p_\perp \mathrm{d}y}(p_\perp^\text{shift}, y^\text{shift})
\end{equation}
\end{widetext}
where the shifted quantities are expressed by
\begin{align}
	\label{eq:RapShift}
	y^\text{shift} &= \operatorname{arcosh}\!\left[\frac{E(p_\perp,y) + \varepsilon}{M_\perp(p_\perp^\text{shift})}\right] - y_\text{beam}\,,\\
	p_\perp^\text{shift} &= \sqrt{{p_\perp^2 + \small{\Delta}p_\perp^2 + 2 p_\perp \small{\Delta}p_\perp \cos \varphi}}\,,\\
	p_\parallel^\text{shift} &= \sqrt{{\big[E(p_\perp,y) + \varepsilon\big]^2 - M_\perp^2(p_\perp^\text{shift})} }\,.
	\label{eq:MomShift}
\end{align}
The convolution integrates over the energy loss $\varepsilon$ and the angle $\varphi$ between the transverse momentum of the bottomonium and the total transverse momentum kick $\small{\Delta}p_\perp$ in the lead nucleus.
It is based on Arleo's and Peign\'{e}'s model of parton energy loss \cite{arleo13,arl13,arl16}, where
partons traversing a medium are expected to loose energy via induced gluon radiation caused by interactions with multiple static scattering centers of the medium.

Hence, the number of produced bottomonia in pPb collisions can be calculated from the production of higher energetic bottomonia in pp collisions and the probability that they will emit the energy difference $\varepsilon$ through gluon radiation.
The probability distribution of an energy loss $\varepsilon$ of bottomonia with energy $E$ in the lead rest frame is given by the normalized quenching weight $P$ \cite{arl13}.
The transverse-momentum kick $\small{\Delta}p_\perp$ is related to the gluon saturation scale in the lead nuclei, and the shifted variables in Eqs.\,(\ref{eq:RapShift}) - (\ref{eq:MomShift}) follow from kinematic considerations.

Further, we also include the gluonic nuclear modification factor $R_g^\text{Pb}$ of the gluon PDF in Pb compared to p in Eq.\,(\ref{eq:AllEffects}).
This is a simplification of the shadowing effects in the Color Evaporation Model as formulated by R.\,Vogt \cite{vogt15}, since we assume that the main contribution to the bottomonium-production cross section comes from gluon fusion.
Hence, the bottomonium momentum fraction $x_2$ is given by the kinematics of $2 \rightarrow 1$ processes,
\begin{align}
	x_2(p_\perp, y) &= \frac{M_{\Y,\perp}}{\sqrtsNN}\exp(-y) \,, \\
	x_2^\text{shift} &= x_2(p_\perp^\text{shift}, y^\text{shift}) \,,
\end{align}
where $M_{\Y,\perp}$ is the transverse mass of the final-state bottomonium.
In our calculations, we use the EPPS16 set \cite{esk17} which includes the most recent global analysis of nuclear shadowing.

Since we consider both, shadowing and coherent energy loss in the cold nuclear matter, we must adapt the value of the transport coefficient $\hat{q}$ that governs the energy-loss model of \cite{arleo13,arl16}, reducing it from $\hat{q} ={}$\SIrange{0.075}{0.046}{\GeV\squared\per\fm\per\clight\squared}.
The difference in the corresponding nuclear modification factors is, however, small because the transverse-momentum kick $\Delta p_\perp$ scales with $\sqrt{\hat{q}}$\,:
The modification is below \SI{3}{\percent} in $p_\perp$-averaged results, and becomes significant only at $p_\perp < \SI{5}{\GeV\per\clight}$.

\subsection{Effective Path Length}
The centrality dependence of the CNM modification factor is caused by the changing effective path length $L_\text{eff}$ which in turn affects the quenching weight $P$.
The path length for a projectile travelling through a medium is proportional to the number of binary collisions and the mean free path.
The latter is given by the inverse of the product of the inelastic pp cross section $\sigma_\text{pp}^\text{inel}$ and the mean number density $\rho_0$ in the nucleus,
\begin{align}
	\label{eq:EffectivePathLength}
	L_\text{eff, Pb}(b) = \frac{\Ncoll(b)}{\rho_0 \ \sigma_\text{pp}^\text{inel}}\,,
\end{align}
with the mean number density 
\begin{align}
	\rho_0 = \frac{208}{V_\text{HS}} = 208\ \frac{3}{4 \pi R_\text{Pb}^3} \approx 0.17\ \text{fm}^{-3}\,,
\end{align}
where $V_\text{HS}$ is the equivalent hard-sphere volume.
Using the results from our Glauber calculation and $\sigma_\text{pp}^\text{inel}(8\, \text{TeV}) \simeq \SI{7.46}{\fm\squared}$ \cite{ant13}, we obtain the value
\begin{align}
	L_\text{eff, Pb}(0) \approx \SI{12.26}{\fm}
\end{align}
in central collisions, which scales with $\Ncoll(b)$ for more peripheral collisions.

Before we proceed to investigate the influence of the hot fireball in asymmetric systems at LHC energies on the $\Y$ suppression, we mention that our calculations for the CNM effects due to both, modifications of the PDFs and coherent energy loss in the nuclear medium (see Sec.\,\ref{sec:results}), are in line with standard results of the CNM-community \cite{alba18}, which were originally presented as predictions before the \SI{8.16}{\TeV} pPb run.

\section{Hot-medium effects}
\label{sec:qgp}
Although the spatial extent of the initial QGP-zone in pPb collisions at LHC energies -- in this work, at $\sqrtsNN = \SI{8.16}{\TeV}$ -- is much less compared to symmetric systems like PbPb, it turns out that the dissociation of bottomonia states in the hot medium is significant and can not be neglected.
We therefore adapt our model for hot-medium bottomonia suppression in symmetric collisions to the case of asymmetric systems.
The bottomonia states are produced with a formation time $\tauF \simeq \SI{0.4}{\fm\per\clight}$ \cite{ngw14,hnw17} in initial hard collisions at finite transverse momentum $p_\perp$, and then move in the hot expanding fireball made of gluons and light quarks where the dissociation processes take place.

\begin{figure}[!tb]
	\centering
	\includegraphics{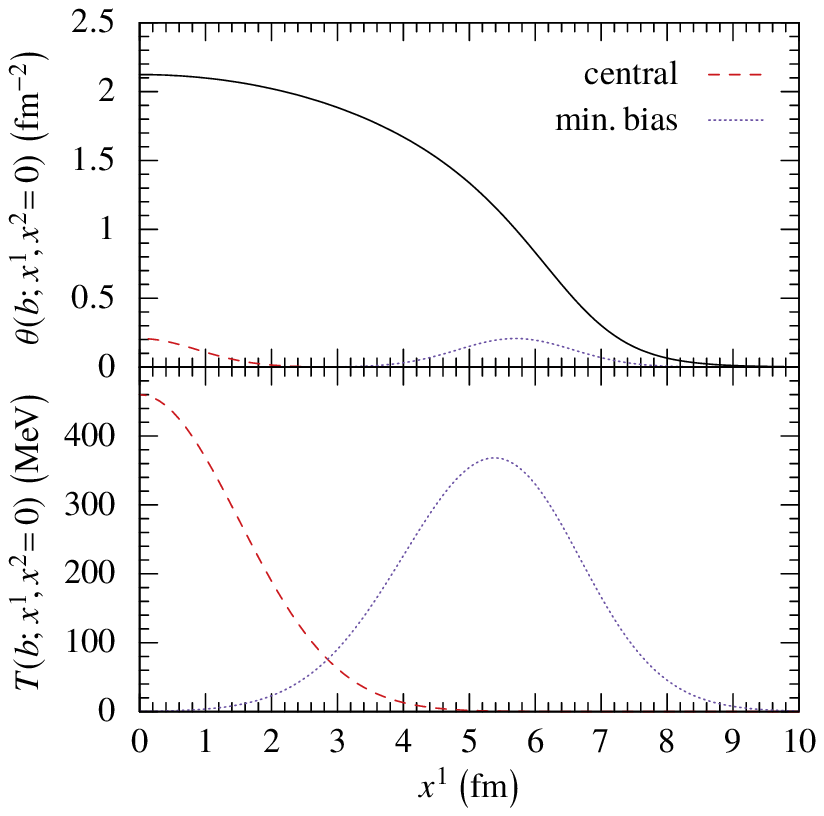}
	\caption{\label{fig2}
		(color online)
		Upper frame: Thickness functions $\theta_\text{Pb}(x^1,x^2{\,=\,}0)$ for Pb (solid curve), and $\theta_\text{p}(x^1,x^2{\,=\,}0)$ for the proton at two different impact parameters, or number of expected binary collisions, corresponding to central collisions ($b=\SI{0}{\fm}$, $\Ncoll \simeq 15.6$, dashed curve) and minimum bias ($b=\SI{5.7}{\fm}$, $\Ncoll \simeq 7$, dotted curve).
		Lower frame: Temperature profiles of the hot QGP generated in pPb collisions at $\sqrtsNN=\SI{8.16}{\TeV}$ for two centralities (dashed, central; dotted, minimum bias) as in the upper frame.
		The initial central temperature is $\Tinit=\SI{460}{\MeV}$.
	}
\end{figure}

\begin{figure*}[!tb]
	\centering
	\includegraphics{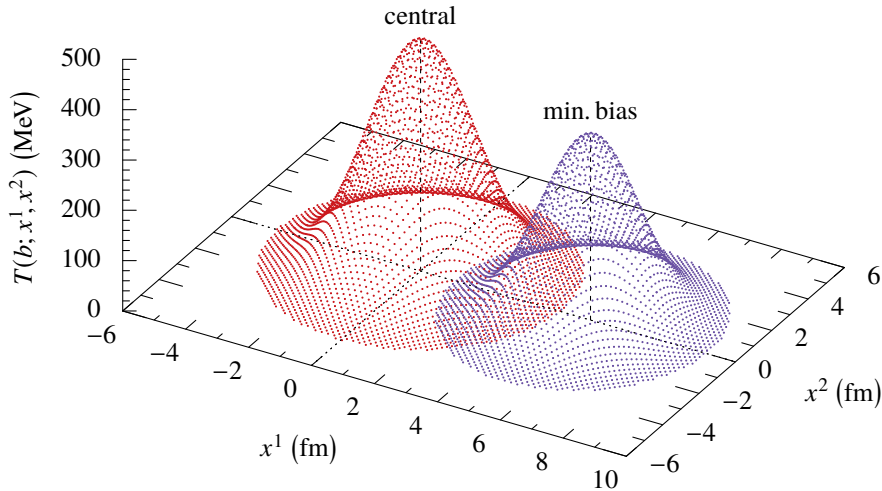}
	\caption{\label{fig3}
		(color online)
		Initial temperature profiles of the hot QGP generated in pPb collisions at $\sqrtsNN=\SI{8.16}{\TeV}$ as functions of the transverse coordinates $(x^1,x^2)$ at two centralities: Central collisions with $\Ncoll \simeq 15.6$, left, and minimum-bias collisions with $\Ncoll \simeq 7$, right, as in Fig.\,\ref{fig2}.
	}
\end{figure*}

The local equilibration time of the fireball is very short, about \SI{0.1}{\fm\per\clight} for gluons \cite{fmr18} and less than \SI{1}{\fm\per\clight} for quarks, such that the conditions for a hydrodynamic treatment of the expansion and cooling of the hot zone are fulfilled.
The difference in the local equilibration time for quarks versus gluons has been discussed on nonequilibrium-statistical grounds in Refs.\,\cite{gw18,gw19res}.
It is essentially due to the role of Pauli's principle, but the different color factors will enhance it.

Indeed the bottomonium formation time in our model is larger than the local equilibration time for gluons, and hence, one may invoke a temperature-dependent formation time, as has been investigated e.g.~by Ko et al.\,\cite{ko15} for heavy symmetric systems.
We had investigated the role of the formation time in our previous work for symmetric systems \cite{ngw13,ngw14}, with the result that there is less suppression for longer formation times, because the system has already cooled, as will be confirmed in Sec.\,\ref{sec:results} for the asymmetric system as well.
This, in turn, requires a higher initial central temperature.

As in case of symmetric systems \cite{ngw14,hnw17}, we use perfect-fluid relativistic hydrodynamics with longitudinal and transverse expansion to account for the background bulk evolution.
The equations of motion are obtained by imposing four-momentum conservation, and solved in the longitudinally co-moving frame (LCF), with metric
\begin{equation}
	\label{eq:metric}
	\begin{gathered}
		g = -\mathrm{d}\tau^2 + (\mathrm{d}x^1)^2 + (\mathrm{d}x^2)^2 + \tau^2 \mathrm{d}y^2,\\
		\tau = \sqrt{(x^0)^2 - (x^3)^2}\,,\qquad
		y = \operatorname{artanh}(x^3/x^0)\,.
	\end{gathered}
\end{equation}
Here, the $x^1$-axis is lying within and the $x^2$-axis orthogonal to the reaction plane, while the $x^3$-axis is parallel to the beam axis.
The resulting equations of motion
\begin{equation}
	\label{eq:eom}
	\partial_\mu (\tau \, T^4 u^\mu u_\nu) = -\tfrac{\tau}{4} \partial_\nu T^4,\qquad
	\partial_\mu (\tau \, T^3 u^\mu) = 0\,,
\end{equation}
for four-velocity~$u$ and temperature distribution~$T$ in the transverse plane $(x^1,x^2)$ are solved numerically, starting at the initial time $\tau_\text{init} = \SI{0.1}{\fm\per\clight}$ in the LCF.
The initial conditions for $u$ and $T$ in symmetric systems are given in Eqs.\,(14)--(16) of Ref.\,\cite{ngw14}.
For the asymmetric system at hand, we adapt the initial condition for $T$ to scale with the distribution of binary collisions in the transverse plane $\ncoll(b;x^1,x^2)$ that lead to the formation of the hot zone,
\begin{equation}
	T(b; \tau_{\text{init}}, x^1, x^2) = \Tinit \, \sqrt[3]{\frac{\ncoll(b;x^1,x^2)}{\ncoll(0;0,0)}}\,.
\end{equation}
In pPb collisions, the expected number of binary collisions in a central collision is $\Ncoll(b{\,=\,}0) \simeq 15.6$, where $\Ncoll(b) = \int\mathrm{d}^2x \, \ncoll(b;x^1,x^2)$.

As discussed in the next section where the comparison with data will be shown, we determine the initial central temperature~$\Tinit$ in pPb collisions at \SI{8.16}{\TeV} by fitting our cold-matter plus hot-medium results to LHCb data at forward rapidities, resulting in $\Tinit=\SI{460}{\MeV}$.
Obtaining instead the initial central temperature from a comparison of hydrodynamic calculations with experimental results for elliptic flow of charged hadrons might be conceivable in the future, once data become available.
It would, however, be less reliable as in case of large symmetric systems, where flow is a more pronounced property.
The inclusion of viscosity would alter our results slightly, allowing for lower temperatures at the same QGP lifetime as compared to perfect-fluid hydrodynamics in our modeling.
It would therefore require a rescaling of the initial central temperature.

The distribution of binary collisions~$\ncoll(b;x^1,x^2)$ is obtained from a Glauber calculation and is proportional to the nuclear overlap function~$\theta_\text{pPb}$,
\begin{gather}
	\theta_\text{pPb}(b;x^1,x^2) = \theta_\text{p}(b;x^1,x^2) \times \theta_\text{Pb}(b;x^1,x^2)\,,\\[2ex]
	\theta_\text{p}(b;x^1,x^2) = \int\mathrm{d}x^3 \rho_\text{p}(|b\vec{e}_1 - \vec{x}|)\,,\\
	\theta_\text{Pb}(b;x^1,x^2) = \int\mathrm{d}x^3 \rho_\text{Pb}(|\vec{x}|)\,,
\end{gather}
where $\rho_\text{p},\rho_\text{Pb}$ are the radial symmetric nucleon distributions of the proton and lead nucleus, respectively.
For the latter, a Woods-Saxon potential with parameters taken from \cite{vries87} is used, whereas we use a Gaussian shape for the proton, with a corresponding radius of \SI{0.875}{\fm}.

The thickness functions $\theta_\text{p},\theta_\text{Pb}$ are displayed in the upper frame of Fig.\,\ref{fig2} for $x^2=0$ and two values of the impact parameter that correspond to central ($b=0$) and ``minimum-bias'' ($\Ncoll(b) = \operatorname{MinBias}\left[\Ncoll\right] \simeq 7$) collisions.
In the lower frame, the corresponding initial temperature profiles along the $x^1$-axis are shown.
The hot medium with $T > \Tcrit \simeq \SI{160}{\MeV}$ is generated once the nuclear overlap of the proton with lead is sufficiently strong.

The full, two-dimensional initial temperature profile in the transverse $(x^1,x^2)$-plane for pPb can be seen in Fig.\,\ref{fig3} for the same impact parameters corresponding to central and minimum-bias collisions as in Fig.\,\ref{fig2}.
Although the hot zone is substantially less extended in pPb as compared to PbPb, it is still sufficiently pronounced to cause in-medium dissociation of the initially produced bottomonia states.

To obtain the hot-medium decay widths of the relevant bottomonia states, the energies $E_{nl}(T)$ and corresponding damping widths $\Gamma^\text{damp}_{nl}(T)$ as a function of QGP temperature~$T$ are needed.
To this end, we solve a radial Schr\"odinger equation with a complex, temperature-dependent potential $V_{nl}(r,T)$ \cite{ngw14} for the six states $\YnS{n}$ and $\chibnP{n}$, $n=1,2,3$, using an iterative method to account for the running of the strong coupling \cite{bethke13}.

Additionally, we derive the width caused by gluon-induced dissociation $\Gamma^\text{diss}_{nl}(T)$~\cite{brezinski-wolschin-2012,ngw14} through an extension of the operator product expansion~\cite{bpes79} and add it incoherently to the damping width.
The two mechanisms emerge in different orders in the effective action, as has been shown in potential nonrelativistic QCD (pNRQCD) approaches~\cite{bram05,bram11}:
The imaginary part of the interaction potential $V_{nl}$ yields collisional damping (``soft process'' in pNRQCD terminology), whereas gluodissociation is described by a singlet-to-octet transition (``ultrasoft process''), and hence, both should be treated individually due to the separation of scales.

Finally, dissociation by screening of the real part of the quark-antiquark potential is taken into account by setting the total decay width to infinity if a state's energy meets the continuum threshold, leading to the total hot-medium decay width \cite{hnw17,hw17}
\begin{equation}
	\Gamma^\text{tot}_{nl} =
	\begin{dcases}
		\Gamma^\text{damp}_{nl} + \Gamma^\text{diss}_{nl} & \text{if}\quad E_{nl} < \lim_{r\to\infty} \operatorname{Re} V_{nl}\,,\\
		\infty & \text{else}\,.
	\end{dcases}
\end{equation}

Due to the high bottom-quark mass, the bottomonia are not expected to be co-moving with the expanding hot medium in the transverse plane.
We consider this finite relative velocity
by applying the relativistic Doppler effect to the medium temperature in the bottomonium rest frame and performing an angular average over the shifted decay widths \cite{hnw17}.



\begin{figure}
	\centering
	\includegraphics{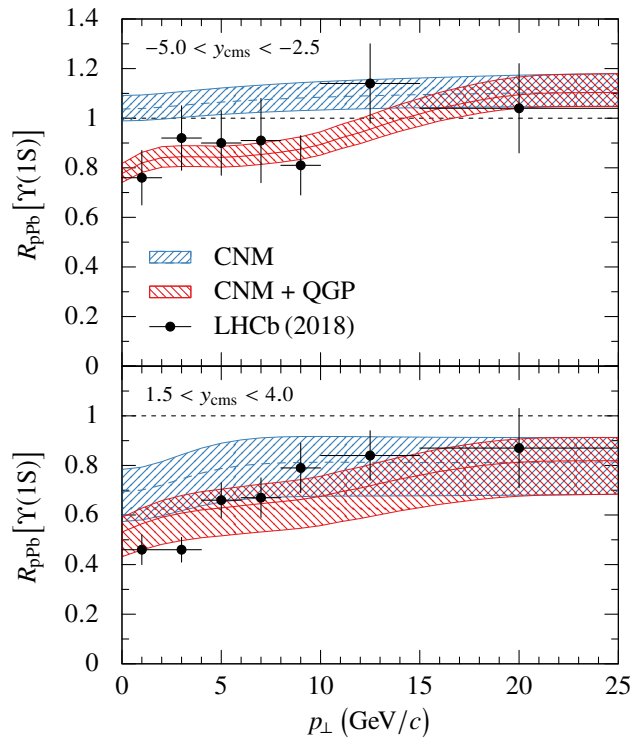}
	\caption{\label{fig4}
		(color online)
		Calculated $p_\perp$-dependent nuclear modification factors $R_\text{pPb}$ for the $\YnS{1}$~spin-triplet ground state in pPb~collisions at $\sqrtsNN = \SI{8.16}{\TeV}$ with LHCb data \cite{lhcb18} in the backward (Pb-going, top) and forward (p-going, bottom) region, for minimum-bias centrality.
		Results for CNM effects that include shadowing, energy loss, and reduced feed-down (dashed curves) are shown together with calculations that incorporate also QGP effects (solid curves).
		The error bands result from the uncertainties of the parton distribution functions that enter the calculations.
	}
\end{figure}

\begin{figure}[b!]
	\centering
	\includegraphics{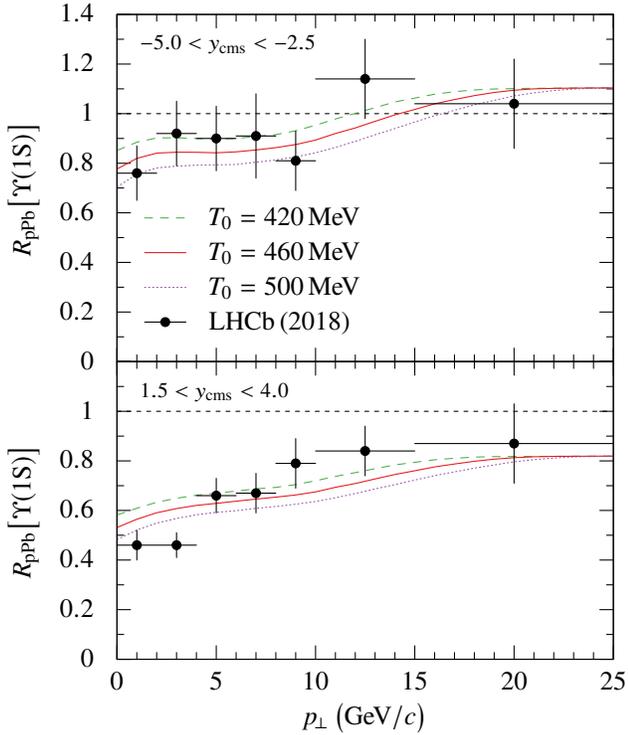}
	\caption{\label{fig5}
		(color online)
		Dependence of the nuclear modification factors $R_\text{pPb}$ for the $\YnS{1}$ state in pPb~collisions at $\sqrtsNN = \SI{8.16}{\TeV}$ on the initial central temperature $\Tinit$ in the backward (Pb-going, top) and forward (p-going, bottom) region, with formation time $\tauF = \SI{0.04}{\fm\per\clight}$.
		Results include CNM and QGP effects.
		Data are from Ref.\,\cite{lhcb18}.
	}
\end{figure}

\begin{figure}[b!]
	\centering
	\includegraphics{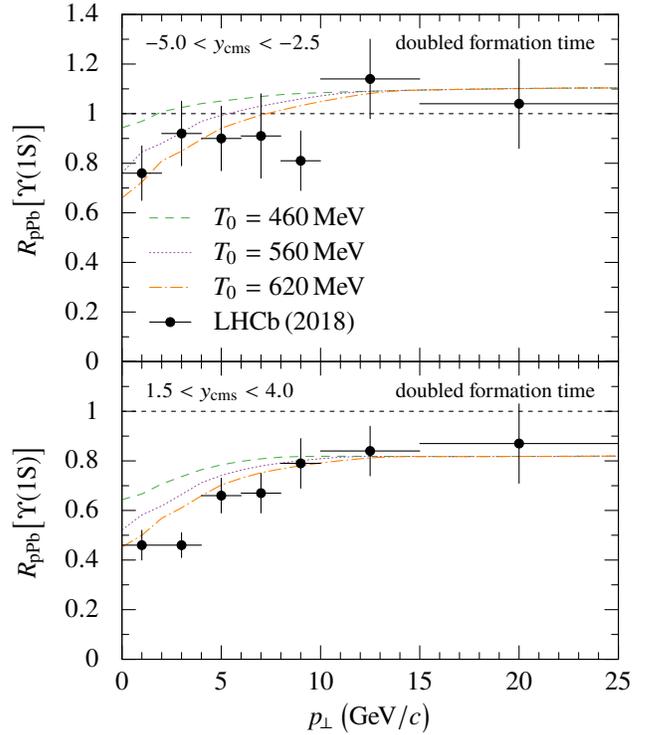}
	\caption{\label{fig6}
		(color online)
		Results for $R_\text{pPb}[\YnS{1}]$ in pPb~collisions at $\sqrtsNN = \SI{8.16}{\TeV}$ when doubling the bottomonia formation time to $\tauF = \SI{0.08}{\fm\per\clight}$ in the backward (Pb-going, top) and forward (p-going, bottom) region, for minimum-bias centrality and three values of the initial central temperature $\Tinit$.
		Data are from Ref.\,\cite{lhcb18}.
	}
\end{figure}
\begin{figure}
	\centering
	\includegraphics{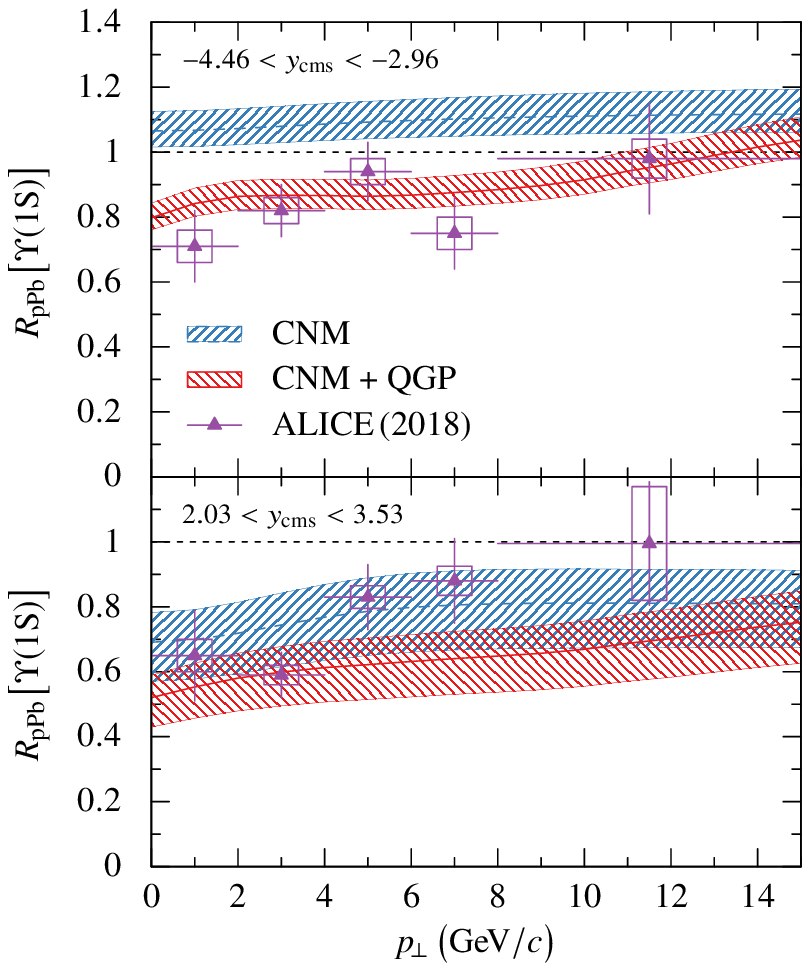}
	\caption{\label{fig7}
		(color online)
		Calculated $p_\perp$-dependent nuclear modification factors $R_\text{pPb}$ for the $\YnS{1}$~state in pPb~collisions at $\sqrtsNN = \SI{8.16}{\TeV}$ with preliminary ALICE data \cite{alice18} in the backward (Pb-going, top) and forward (p-going, bottom) region.
		The rapidity regions are slightly different from Fig.\,\ref{fig4}.
		Results for CNM effects that include shadowing, energy loss, and reduced feed-down (dashed curves) are shown together with calculations that incorporate also QGP effects (solid curves).
		The error bands result from the uncertainties of the parton distribution functions that enter the calculations.
	}
\end{figure}
\begin{figure}
	\centering
	\includegraphics{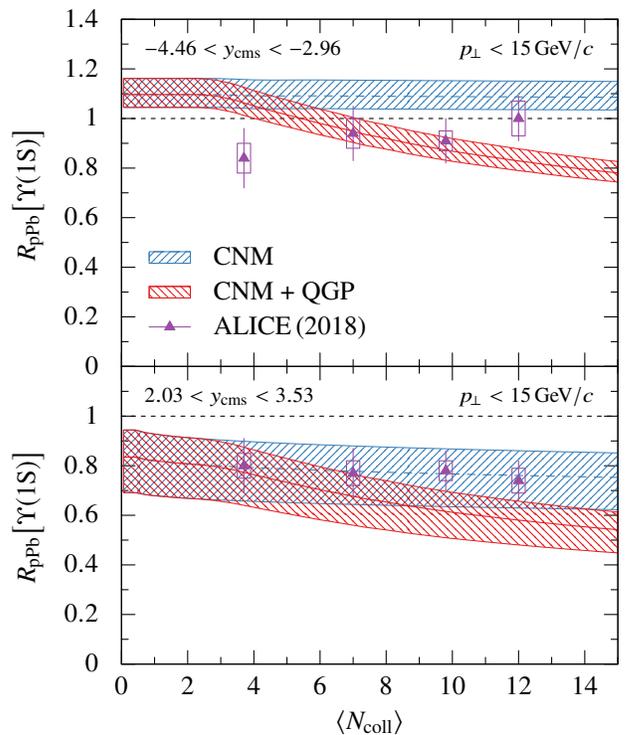}
	\caption{\label{fig8}
		(color online)
		Calculated centrality-dependent nuclear modification factors $R_\text{pPb}$ for the $\YnS{1}$~state in pPb~collisions at $\sqrtsNN = \SI{8.16}{\TeV}$ with preliminary ALICE data \cite{alice18} in the backward (Pb-going, top) and forward (p-going, bottom) region.
		The rapidity regions are as in Fig.\,\ref{fig7}.
		Results for CNM effects that include shadowing, energy loss, and reduced feed-down (dashed curves) are shown together with calculations that incorporate also QGP effects (solid curves).
		The error bands result from the uncertainties of the parton distribution functions that enter the calculations.
	}
\end{figure}

\begin{figure}
	\centering
	\includegraphics{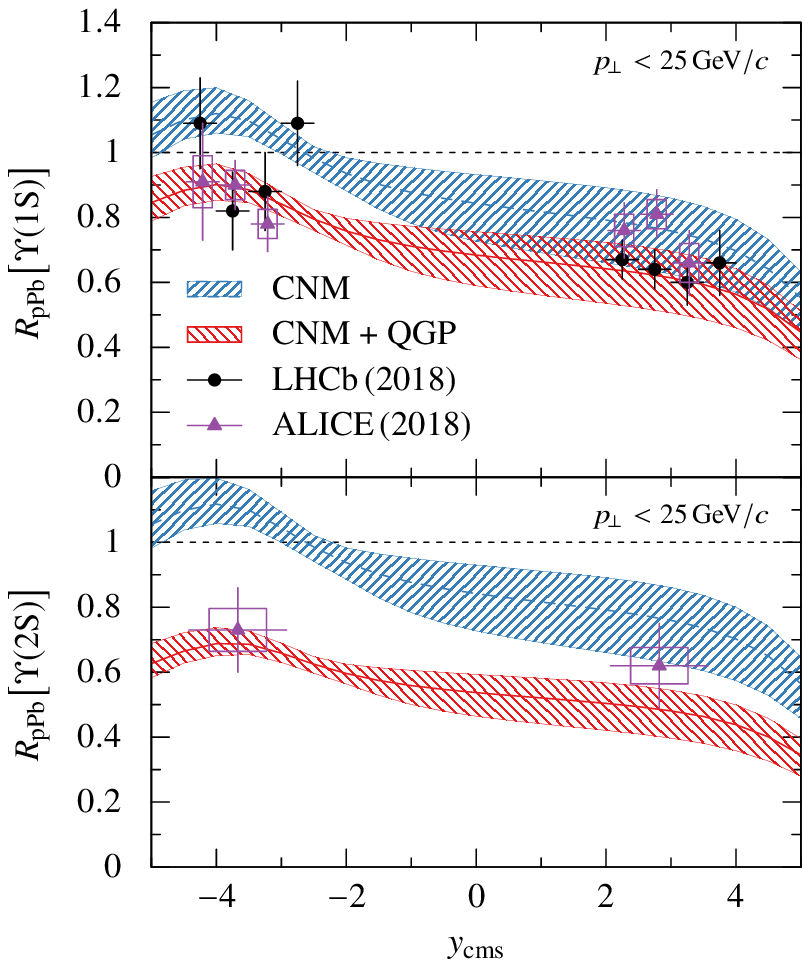}
	\caption{\label{fig9}
		(color online)
		Calculated rapidity-dependent nuclear modification factors $R_\text{pPb}$ for the $\YnS{1}$ (top) and $\YnS{2}$~state (bottom) in pPb~collisions at $\sqrtsNN = \SI{8.16}{\TeV}$ with preliminary ALICE data \cite{alice18} (triangles) and with LHCb data \cite{lhcb18} (circles).
		Results for CNM effects that include shadowing, energy loss, and reduced feed-down (dashed curves) are shown together with calculations that incorporate also QGP effects (solid curves).
		The error bands result from the uncertainties of the parton distribution functions that enter the calculations.
	}
\end{figure}

\section{Comparison to Data}
\label{sec:results}
We investigate the impact of cold-matter and hot-medium effects on bottomonium yields in pPb collisions, and compare them to the most recent results provided by the CERN collaborations LHCb and ALICE.
We calculate the nuclear modification factors for bottomonium in pPb collisions as follows:
First, we apply the inverse bottomonium-decay cascade to the corresponding pp cross sections to obtain the cross sections for bottomonium production in pp.
The latter are then modified for pPb with the CNM effects as discussed in Sec.\,\ref{sec:cnm}, as well as with the thermal QGP effects in the hot zone from Sec.\,\ref{sec:qgp}.
Finally, these are propagated through the bottomonium-decay cascade which yields the cross sections for bottomonium decays in pPb collisions.
In comparisons with transverse-momentum-, centrality-, and rapidity-dependent data, the respective other variables are integrated over centralities or the corresponding kinematical regions according to Eqs.\,(\ref{eq:cFactor}) and (\ref{eq:pTFactor}).

Our model results for the $\YnS{1}$ ground state are compared with the transverse-momentum dependent \SI{8.16}{\TeV} pPb data from the LHCb collaboration \cite{lhcb18} in Fig.\,\ref{fig4}: Upper frame in the backward (Pb-going) region with $-5<y_\text{cms}<-2.5$, lower frame in the forward (p-going) region with $1.5<y_\text{cms}<4$.
In both cases, calculations for cold-matter effects as obtained from initial-state PDF modifications plus energy loss are displayed in the upper bands, whereas the lower bands include the hot-medium suppression.
The small additional suppression from reduced feed-down is included in both cases; if only CNM effects are considered, its impact is negligible.
The broad bands refer to the uncertainties of the parton distribution functions only, not to additional systematic uncertainties that may arise from inherent model simplifications or uncertainties in the choice of parameters.

In the backward region, the additional hot-medium suppression beyond the CNM effects at low transverse momentum clearly improves the agreement with the LHCb data.
Here, the CNM effects alone predict an enhancement of the nuclear modification factor above one due to antishadowing, whereas the LHCb data in the region $p_\perp<\SI{10}{\GeV\per\clight}$ show suppression, which is well reproduced by our hot-medium plus cold-matter results.

The plateau that the data display at $p_\perp<\SI{10}{\GeV\per\clight}$ is a consequence of the relativistic Doppler effect due to the velocity of the moving bottomonia relative to the expanding QGP in our hot-medium model \cite{ngw14,hnw17}:
With rising $p_\perp$, the anisotropic effective temperature that the bottomonia experience is Doppler-shifted.
As a consequence, the angular-averaged values of the corresponding decay widths cause a flat $p_\perp$-dependence of $R_\text{pPb}$ \cite{hnw17}.
At $p_\perp>\SI{10}{\GeV\per\clight}$, the CNM effects are, however, more important than the QGP suppression.
In the forward region, the additional hot-medium suppression at $p_\perp<\SI{10}{\GeV\per\clight}$ also improves the agreement with the LHCb data substantially.

The value of the initial central temperature $\Tinit = \SI{460}{\MeV}$ that we use in our calculation for the hot zone together with a bottomonia formation time of $\tauF = \SI{0.4}{\fm\per\clight}$ results in reasonable agreement with the LHCb data, but it is obviously model-dependent because a larger formation time would allow for more cooling, thus requiring a larger initial temperature.
Hence, we first investigate the dependence of the system on the initial central temperature $\Tinit$, keeping the formation time constant.
Typical results are shown in Fig.\,\ref{fig5}, where curves for $\Tinit = {}$\SIlist{420;460;500}{\MeV} are displayed, each of them showing substantially different suppression as expected.

Whereas the results in Fig.\,\ref{fig5} refer to a formation time $\tauF = \SI{0.4}{\fm\per\clight}$ \cite{ngw14,hnw17}, in Fig.\,\ref{fig6} we investigate the effect of doubling the formation time of all six bottomonia states to $\tauF = \SI{0.8}{\fm\per\clight}$.
This causes less suppression in both, backward and forward direction, and can be cured by choosing a larger initial central temperature around $\Tinit \simeq \SI{600}{\MeV}$.
The investigation of the correlation between $\Tinit$ and $\tauF$ should in the future be supplemented by a more detailed model for the temperature-dependence of the formation time, as has already been done in Ref.\,\cite{ko15} at energies available at the Relativistic Heavy Ion Collider.
For the present investigation, however, we keep the formation time constant, and adapt the initial central temperature to be $\Tinit = \SI{460}{\MeV}$.

The comparison of our transverse-momentum dependent results with preliminary ALICE data \cite{alice18} in Fig.\,\ref{fig7} shows then also agreement in the backward region (top) for a slightly different rapidity band, $-4.46<y_\text{cms}<-2.96$.
In the forward region with $2.03<y_\text{cms}<3.53$, however, the comparison between our results and the ALICE preliminary data is less convincing than in the LHCb case.

For the centrality dependence displayed in Fig.\,\ref{fig8}, the CNM effects result in a fairly flat dependence on the number of binary collisions both backward -- where antishadowing enhances $R_\text{pPb}$ above one -- and forward, where shadowing and energy loss already cause suppression.
The hot-medium contributions generate even more suppression in central collisions.
This disagrees with the preliminary ALICE data which show almost no centrality dependence in the forward region, and backwards even a slight rise of $R_\text{pPb}$ with increasing centrality.
The origin of the discrepancy is presently an open question.
Note that ALICE data for $J/\psi$ modification factors in \SI{5.02}{\TeV} pPb collisions show an even stronger rise towards $R_\text{pPb}\simeq 1.2$ with increasing centrality in the backward region \cite{adam15}, although there is growing suppression in the forward region.

Regarding the rapidity dependence of the $\YnS{1}$ modification factor displayed in Fig.\,\ref{fig9}, the characteristic forward-backward asymmetric shape that is caused by the CNM effects -- with $R_\text{pPb}>1$ in the backward region due to antishadowing, but $R_\text{pPb}<1$ in the forward region due to shadowing and energy loss -- is maintained, but smoothened once the hot-medium effects are added.
In particular, these cause an overall suppression of $R_\text{pPb}$ below one even in the backward region, thus improving the agreement with LHCb and preliminary ALICE data in this region.

The more pronounced suppression of the excited $\YnS{2}$ state as compared to the ground state that ALICE has found in both forward and backward regions as shown in the lower frame of Fig.\,\ref{fig9} can not result from CNM effects: These yield enhancement above one, rather than suppression in the backward region.
The predicted $\YnS{2}$ CNM enhancement is in magnitude quite similar to the one of the ground state, whereas the preliminary ALICE data show suppression down to almost \SI{70}{\percent}, in reasonable agreement with our calculation.
This result strongly underlines the importance of hot-medium effects in the observed bottomonia suppression in pPb collisions at LHC energies.

In all our calculations, the formation time for the six included bottomonia states is \SI{0.4}{\fm\per\clight} and the initial central temperature of the fireball is $\Tinit=\SI{460}{\MeV}$ -- which is somewhat less than the initial central temperature of \SI{480}{\MeV} in \SI{2.76}{\TeV} PbPb collisions or the extrapolated value of \SI{513}{\MeV} in \SI{5.02}{\TeV} PbPb collisions that resulted in agreement of our corresponding predictions with CMS data \cite{gw19}.
Modifications due to a different -- and possibly, state-dependent -- value of the formation time had been discussed in case of symmetric systems in Ref.\,\cite{ngw14}.
The initial central temperature has been determined from the difference between standard CNM-results and data of the LHCb and ALICE collaborations.
It will be interesting to see if future determinations of the initial central temperature from other observables yield similar values.

\section{Conclusion}
\label{sec:conclusion}
In summary, we have investigated the modifications of $\Y$ yields in pPb~collisions at a LHC energy of \SI{8.16}{\TeV} in relation to scaled pp collisions as functions of transverse momentum, rapidity, and centrality due to both CNM and QGP effects.

We have considered in the initial stages of the collision the CNM effects of shadowing and antishadowing due to the modifications of the parton distribution functions, for which we take most recent values.
The partonic energy loss has been accounted for within an established model for parton propagation from the initial to the final state.
As a new development, we have combined this well-known CNM treatment with our model for bottomonia suppression through hot-medium effects.
So far, it has only been applied to symmetric systems such as PbPb, where its predictions were found to agree with the measured $\YnS{1}$ suppression.
In the hot zone, rapid initial local equilibration of quarks and even faster equilibration of gluons ensures that a hydrodynamic approach is applicable, and we consider the corresponding longitudinal, but also transverse expansion, in a perfect-fluid model.
In the course of the expansion and cooling, we explicitly treat the hot-medium processes gluodissociation, screening and damping, until the temperature falls below the critical value.

In the asymmetric pPb system, the hot-medium suppression turns out to be quite relevant, in spite of the spatially less extended hot zone as compared to symmetric systems.
The feed-down cascade from the excited bottomonia states produces some additional ground-state suppression due to melting or depopulation of the excited states, but this contribution is not as significant as in PbPb at \SI{5.02}{\TeV} where the excited states are almost totally screened or depopulated and therefore the feed-down to the ground state is substantially reduced.

Summarizing our comparisons with recent LHC data on pPb collisions at \SI{8.16}{\TeV}, we conclude that not only cold nuclear matter effects but also $\Y$ suppression in the hot medium are responsible for the observed modifications of $\Y$ yields in pPb collisions as compared to pp.
The hot-medium effects are well represented by our model that has shown considerable predictive properties for symmetric systems, and is evidently also well-suited for smaller and asymmetric systems at energies reached at the Large Hadron Collider, where a less extended fireball of hot quark-gluon plasma is created.

\begin{acknowledgments}
	J.\,H.~acknowledges a PhD fellowship of the Villigst Scholarship Foundation.
	This work receives Open Access funding by SCOAP$^3$.
\end{acknowledgments}
\newpage



\bibliography{gw19}





\end{document}